\documentclass[11pt]{article}       
 \usepackage{amsmath}
\usepackage{graphicx}
\usepackage{amsfonts}
\usepackage{authblk}

\newcommand{\figscale}{0.6}
\begin{document}

\title{Anchor points for genome alignment based on 
{\em Filtered Spaced Word Matches}}

\author[1]{Chris-Andr\'{e} Leimeister}
\author[1]{Thomas Dencker} 
\author[1,2]{Burkhard Morgenstern}

\affil[1]{ University of G\"ottingen, Department of Bioinformatics, 
Goldschmidtstr. 1, 37077 G\"ottingen, Germany}  

\affil[2]{University of G\"ottingen, Center for Computational Sciences,  
Goldschmidtstr. 7, 37077 G\"ottingen, Germany}


\maketitle

\begin{abstract} 

Alignment of large genomic sequences is a fundamental task in computational
genome analysis. Most methods for genomic alignment use
high-scoring local alignments as  {\em anchor points} to reduce the search
space of the alignment procedure. Speed and quality
of these methods therefore depend on the underlying anchor points.
Herein, we propose to use {\em Filtered Spaced Word Matches}
to calculate anchor points for genome alignment. To evaluate this approach,
we used these anchor points in the the widely
used alignment pipeline {\em Mugsy}. For distantly related
sequence sets, we could substantially
improve the quality of alignments produced by {\em Mugsy}. 

\end{abstract}

\section*{Introduction}
Sequence comparison is one of the most fundamental tasks in computational biology. Here, a basic task is to {\em align} two or several DNA or protein sequences -- either {\em globally}, over their entire length, or {\em locally}, by
restricting the alignment to a single region of homology. 
Standard approaches to sequence alignment assume that the input sequences
derived from a common ancestral sequence, and that evolutionary events  
are limited to substitutions, insertions and deletions of single residues
or small sequence segments. In this case, sequence homologies can be 
represented by {\em global sequence alignments}, that is by inserting 
{\em gap characters} into the sequences such that 
evolutionarily related sequence positions are arranged on top of each
other.  
Under most scoring schemes, calculating an \textit{optimal} alignment of two sequences takes time proportional to the product of their lengths and is therefore limited to rather short sequences \cite{nee:wun:70,smi:wat:81a,got:82,mor:02,dur:edd:kro:mit:98}. 

With the rapidly increasing number of partially or fully sequenced genomes, 
alignment of {\em genomic} sequences has become an important field of research in bioinformatics, see \cite{ear:ngu:hic:etal:14} for a recent review and evaluation of some
of the most popular approaches. 
Here, the first challenge is the sheer size of the input sequence that makes
it impossible to use traditional algorithms with quadratic run time.  
The second challenge is that related genomes often share {\em multiple} 
regions
of local sequence homology, interrupted by non-conserved parts of the sequence
where no significant similarities can be detected. This means that neither
{\em global} nor {\em local} alignment methods can properly represent 
the homologies between whole genomes. Finally, evolutionary 
events such as duplications and large-scale rearrangements must be taken into 
account. 
Since it is not possible, in general, to represent homologies 
among genomes in one single alignment, advanced genome aligners 
return alignments of so-called {\em Locally Collinear Blocks}, {\em i.e.}
blocks of  
segments of the input sequences that contain the same 
genes in the same relative order. 

Since the late Nineteen Nineties, major efforts have been made to a
address the problem of genome alignment, and many approaches have been 
published. 
One of the first multiple-alignment programs that was applied to genomic 
sequences was {\em DIALIGN} \cite{mor:dre:wer:96,mor:rin:abd:etal:02}. 
This program 
composes multiple alignments from chains of local pairwise 
alignments, and it does not penalize  gaps; it is therefore
able to align sequences where local homologies are separated by long 
non-homologous segments. The program has been applied, for example,
 to identify small non-coding functional elements in genomic sequences 
\cite{bar:goe:ger:etal:01,cha:cha:kin:etal:03}.  
However, the program was initially not designed for large genomic sequences, 
and it is limited to sequences up to around 10~{\em kb}. 
Moreover, {\em DIALIGN} is not able to deal with duplications, 
rearrangements and homologies on inverse strands of genomes.

To align longer sequences, most programs for genomic alignment 
rely on some sort of {\em anchoring} 
\cite{hua:umb:lep:06,mor:pro:poe:sta:06},
In a first step, they use a fast method for local alignment to identify
high-scoring  local homologies, so-called {\em anchor points}. 
Next, {\em chains} of such local alignments are calculated and, finally,
sequence segments between the chained high-scoring local alignments  
are aligned with a slower but more sensitive alignment method. 
For multiple sequence sets, anchor points can be defined either between 
pairs of sequences or between several or all of the input
sequences. 
A pioneering tool to find anchor points for genomic alignment was 
 {\em MUMmer} \cite{del:kas:fle:99}; the current version of the program
\cite{kur:phi:del:etal:04} is considered the state-of-the-art 
in alignment anchoring. 
MUMmer
 uses \textit{maximal unique matches} as pairwise anchor points to align 
genomic sequences or protein sequences. By contrast, {\em MGA} \cite{hoe:kur:ohl:02} is a tool for {\em multiple} alignment of genomic sequences that uses \textit{maximal exact matches}  
between {\em all} sequences within a given sequence set.
Both {\em MUMmer} and {\em MGA} use {\em suffix trees} \cite{kur:99} to 
rapidly identify pairs or blocks of identical words, 
one word from each of the
sequences, that are then used as anchor points.  
Both programs are able to align entire bacterial genomes,
{\em MUMmer} was also used in the {\em  A. thaliana} genome project 
\cite{agi:00}. However, since 
the probability of homologous exact matches rapidly decreases with increasing 
divergence, they are most useful to compare closely related
genomes, such as different strains of {\em E. coli}.  

Other approaches to genome alignment are {\em OWEN} \cite{ogu:roy:sha:kon:02}, 
\textit{AVID} \cite{bra:dub:pac:03}, {\em MAVID} \cite{bra:pac:03}, 
 \textit{LAGAN and Multi-LAGAN} \cite{bru:do:coo:etal:03}, 
{\em CHAOS/DIALIGN} \cite{bru:cha:goe:etal:03l},   
the \textit{VISTA genome pipeline} \cite{dub:pol:kis:bru:09}, 
\textit{TBA} \cite{bla:ken:rie:etal:04} and {\em Mauve} 
\cite{dar:mau:bla:per:04}, see \cite{dew:pac:06,bat:05} for a review. 
All of these methods are based
on alignment anchoring, and most of them are able to deal with duplications
and genome rearrangements. 
Some methods for genomic alignments are based on {\em statistical} properties 
of the sequences \cite{bra:rob:smo:etal:09,dar:mau:bla:per:04}.  
Other methods are based on {\em graphs}, for example on 
\textit{A-Bruijn graphs} \cite{rap:zhi:tan:pev:04} or on 
\textit{cactus graphs} \cite{pat:ear:ngu:etal:11}. 
A further development of {\em Mauve} called
{\em progressiveMauve} uses palindromic {\em spaced seeds}
instead of exact word matches as anchor points  \cite{dar:mau:per:10}.  
That is, for a given binary pattern of length $\ell$ representing 
{\em match} and {\em don't-care} positions, one searches for a set of  
$\ell$-mers, one $\ell$-mer from each of the input sequences, such that
all $\ell$-mers have matching nucleotides  
at the {\em match} positions. At the {\em don't-care}
positions, mismatches are allowed. Palindromic patterns are used to cover
both strands of the input sequences.  Spaced seeds are 
used in database searching
\cite{ma:tro:li:02,dar:tre:zha:etal:06} and alignment-free 
sequence comparison \cite{lei:bod:hor:lin:mor:14} since they have been
shown to lead to better results than contiguous word matches.

\textit{Mugsy} \cite{ang:sal:11} is a popular software pipeline for multiple 
whole-genome alignment. 
In  a first step, this program uses \textit{Nucmer} \cite{kur:phi:del:etal:04} 
to construct all pairwise alignments of the input sequences. {\em Nucmer},
in turn,  
uses {\em MUMmer} to find exact unique word matches which are used as 
alignment anchor points.  An 
{\em alignment graph} is constructed from these pairwise alignments
using the {\em SeqAn} software \cite{doe:wee:rau:rei:08}, and 
{\em Locally Collinear Blocks} are constructed.  
Finally, a multiple alignment is calculated using {\em SeqAn::TCoffee} 
 \cite{rau:emd:wee:etal:08}. {\em Mugsy} has been designed 
to align closely related genomes, such as different strains of a 
bacterium. Here, it produces alignments of high
quality. On more distantly related genomes, however, the program is often 
outperformed by other multiple genome aligners \cite{ear:ngu:hic:etal:14}.

Finding {\em anchor points} is the most important step in whole-genome sequence alignment. Here, a trade-off between {\em speed},  
{\em sensitivity} and {\em precision} is necessary.  
A sufficient number of anchor points is required in order
to reduce the search space and thereby the run time for the subsequent, 
more sensitive alignment routine.  Wrongly chosen anchor points, 
on the other hand, can substantially deteriorate the
quality of the final output alignment. If spurious similarities are used
as anchor points, this not only results in non-homologous parts of the
sequences being aligned. Wrong anchor points may also prevent the program 
from aligning biologically relevant, true homologies since aligning them
may be incompatible with the selected anchors. 
Also, if the number of anchor points is too large, finding optimal  
chains of anchor points can become computationally expensive.

In this paper, we propose a novel algorithm to find pairwise anchor points
for genomic alignments that is based on the {\em Filtered Spaced Word
Matches (FSWM)} idea that we previously introduced \cite{lei:soh:mor:17}. 
Anchor points are calculated using a {\em hit-and-extend} approach 
where high-scoring spaced-word matches are used as {\em seeds}:
for an underlying binary pattern of length~$\ell$ representing 
{\em match} and {\em don't care}
positions, we rapidly identify {\em spaced-word matches}, {\em i.e.} 
length-$\ell$ segment pairs from the input
sequences with matching nucleotides at the {\em match} positions
but with possible mismatches at the {\em don't care} positions.    
For each spaced-word match, we then calculate a similarity score considering 
{\em all} aligned positions -- including the {\em don't-care positions} --,  
and we keep only those spaced-word matches that have a score above a certain
threshold.  These segment pairs are then extended to 
locally-maximal gap-free alignments, similar as in {\em BLAST}
\cite{alt:etal:90}. 
To evaluate our anchoring approach, we used the 
{\em Mugsy} pipeline using our software in the initial step, to find
anchor points.  
For closely related input
sequences, the quality of the resulting alignments is comparable to the
original version of {\em Mugsy} where exact word matches are used for 
anchoring.  
Our paproach is far superior, however,
if distal sequences are to be aligned, where most other alignment 
approaches either fail to produce alignments or require an unacceptable 
amount of time.

Through our web site, we
provide the adapted {\em Mugsy} pipeline with our anchoring approach
as a pipeline for genome-sequence alignment that can be readily installed.  
A standalone version of our {\em spaced-words} software is provided as well, 
such that developers can integrate it into their own sequence-analysis
pipelines.

\section*{Filtered Spaced Word Matches}
 For a sequence $S$ of length~$L$ over an alphabet $\Sigma$ 
and $0 < i \le L$, $S[i]$ denotes the $i$-th symbol of~$S$. 
For integers $w\le \ell$, 
a binary {\em pattern}~$P$  of length~$\ell$ and weight~$w$ is a 
word over $\{0,1\}$ of length $\ell$ such that there are exactly
$w$ indices~$i$ with $P[i]=1$. These positions  are
called {\em match positions}, while positions~$i$ with $P[i]=0$ are
called {\em don't-care positions}. 
A {\em spaced word} with respect to a pattern~$P$ is a word~$w$ over
$\Sigma \cup \{\ast\}$ where `$\ast$' is a wildcard character  not contained
in $\Sigma$, and $w[k]=\ast$ holds if and only if $k$ is a {\em don't-care
position}, i.e. if $P[k]=0$, see also 
\cite{lei:bod:hor:lin:mor:14,hor:lin:bod:etal:14}.
A spaced word $w$ with respect to a pattern~$P$ occurs in a sequence
$S$ at position $i$ if $S[i+k-1] = w[k]$ for all match positions $k$
of the pattern~$P$. 

For sequences $S_1$ and $S_2$ with lengths $L_1$ and $L_2$, respectively
and  a  pattern~$P$ of length $\ell$, and 
$1\le i \le L_1-\ell+1, 1\le j \le L_2-\ell+1$, 
 we say that there is  {\em spaced-word match} between $S_1$ and $S_2$ 
at $(i,j)$ with respect to~$P$ if 
the spaced words at $i$ in $S_1$ and at $j$ in $S_2$ are identical - in
other words, if 
for all match positions $k$ in $P$, one has  
\[ S_1[i+k-1] = S_2[j+k-1].\] 
Below is a spaced-word match 
between two {\em DNA} sequences $S_1$ and $S_2$ at $(5,2)$  
with respect  to the  pattern $P=1100101$: 
\[
\begin{array}{lcccccccccccccc}
S_1: & G & C & T & G & T & A & T & A & C & G & T & C &   \\
S_2: &   &   &   & S & T & A & C & A & C & T & T & A & T \\  
P:   &   &   &   &   & 1 & 1 & 0 & 0 & 1 & 0 & 1 &   &   \\
\end{array}
\]
Indeed, the spaced word `$TA\ast\ast C\ast T$' occurs at positions $5$ in $S_1$ 
and at position $2$ in $S_2$. 

Herein, we propose to use spaced-word matches as a first step to calculate 
{\em anchor points} for pairwise alignment. 
We therefore need some criterion to distinguish between spaced-word matches
representing {\em true homologies} and random {\em background} matches. 
In a previous paper, we used spaced-word matches to estimate 
phylogenetic distances between genomic sequences \cite{lei:soh:mor:17}. 
To this end, we first identified
all spaced-word matches with respect to a given pattern~$P$. To remove spurious
random spaced-word matches, we applied a simple {\em filtering procedure}: 
using a nucleotide substitution matrix \cite{chi:yap:mil:02},
we calculated for each spaced-word match the sum scores of all aligned
pairs of nucleotides (including match and don't-care positions), and
we removed all spaced-word matches with a score below zero.  

\begin{figure*}[t]
\begin{center}
\includegraphics[width=14cm]{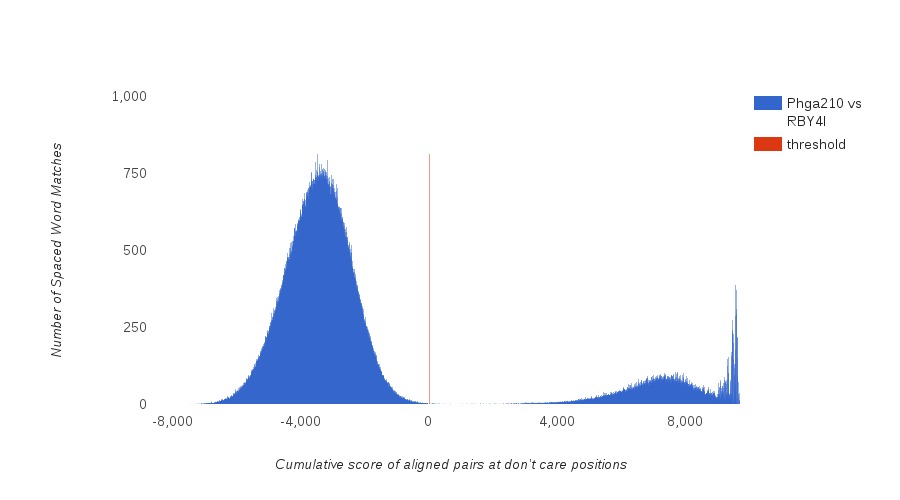}
\end{center}
\caption{\label{fig_sw_histo} {\em Spaced-words histogram} for a comparison of
two bacterial genomes, 
{\em Phaeobacter gallaeciensis} 2.10 and {\em Rhodobacterales bacterium Y4I}.
 All possible spaced-word matches with respect to a given
binary pattern~$P$ are identified, and their scores are calculated
as explained in the main text. The number of spaced-word matches with
a score $s$ is plotted against $s$. Two peaks are visible, an approximately
normally distributed 
peak for background spaced-word matches, and a more complex 
peak for spaced-word matches representing homologies. With a cut-off
value of zero, background and homologous spaced-word matches can be
reliably separated. 
}  
\end{figure*}

A graphical representation of the spaced-word matches between two 
sequences shows that 
this procedure can clearly separate random spaced-word matches from 
true homologies. If we plot for each possible score value~$s$ the number of
spaced-word matches with score $s$, we obtain a bimodal distribution
with one peak for random  matches and  a second peak for homologies. 
We call such a plot a {\em spaced-words histogram}. 
For simulated sequence pairs under a simple model of evolution, both peaks
are normally distributed. For real-world sequences, the random peak is
still normally distributed, but the `homologous' peak is more complex, 
see Figure~\ref{fig_sw_histo}. Even so, using a cut-off value of zero can clearly
distinguish between random matches and true homologies.  
More examples for {\em spaced-words histograms}  are given in 
\cite{lei:soh:mor:17}.

Our approach to find anchor points for pairwise genomic alignment is as 
follows. For given parameters $\ell$ and $w$, we first
calculate a binary pattern with length $\ell$ and weight (number
of {\em match} positions) $w$ using our recently developed 
software {\em rasbhari} \cite{hah:lei:oun:etal:16}. 
We then identify all spaced-word matches with respect to~$P$. 
To find homologies even for distantly related sequences, 
we use patterns with a low weight; by default, we use
a weight of $w=10$. 
On the other hand, we use a large number of  {\em don't-care}
positions, since this makes it easier to distinguish true homologies
from random spaced-word matches. 
By default, we use a pattern length of $\ell = 110$, so our patterns
contain 10 match positions and 100 don't-care positions; 
we use the following nucleotide substitution matrix described in  
 \cite{chi:yap:mil:02}:
\[
\begin{array}{crrrr}
  & A  &   C  &  G  &   T  \\
A & 91 & -114 & -31 & -123 \\
C &    & 100  &-125 &  -31 \\
G &    &      & 100 & -114 \\
T &    &      &     &   91 \\
\end{array}
\]
Based on this matrix, we calculate the {\em score} of each spaced-word
match as the sum of the substitution scores of all aligned pairs of nucleotides.
We then discard all spaced-word matches with a score below zero. 

Next, we extend the identified spaced-word matches in both 
directions without gaps. As the starting point for this extension,
we do not use the full spaced-word matches, but their mid points. 
The reason for this is that, with our long patterns, even a high-scoring
spaced-word match may not represent sequence homologies over its 
entire length. It often occurs that some part of a spaced-word
alignes homologous nucleotides, but another part extends into non-homologous
regions of the sequences. There is a high probability, however, that
the mid point of a long, high-scoring spaced-word match is located within
a region of true homology.  
Finally, we use the produced `extended' gap-free alignments as 
anchor points for alignment.

\section*{Evaluation}
To evaluate {\em Filtered Spaced Word Matches (FSWM)} and to compare it to the 
state-of-the-art
approach to alignment anchoring, we used the  
{\em Mugsy} software system. As mentioned above, the original {\em Mugsy}
uses  {\em MUMmer} to find pairwise anchor points. We replaced {\em MUMmer}
in the {\em Mugsy} pipeline 
by our {\em FSWM}-based anchor points  and evaluated the resulting multiple alignments.  
In addition, we compared these alignments to alignments produced by 
the multiple genome aligner {\em Cactus} \cite{pat:ear:ngu:etal:11}. 
{\em Cactus} is known to be one of the best existing tools for multiple genome 
alignment; it performed excellent in the 
{\em Alignathon} study \cite{ear:ngu:hic:etal:14}.  
%
%
%
To measure the performance of the compared methods, we used simulated 
genomic sequences as well as three sets of real genomes. 
%
To make {\em MUMmer} directly comparable to {\em FSWM}, we used a minimum 
length of 10 {\em nt} for 
maximum unique matches, corresponding to the default {\em weight} (sum 
of {\em match positions}) used in {\em Spaced Words}.  
 Note that, by default, {\em MUMmer}
uses a minimum length of 15 {\em nt}. With this default value, however,
we obtained alignments of much lower quality. 
{The \textit{Cactus} tool was run with default values.}

 \subsection*{Simulated genome sequences }
To simulate genomic sequences, we used the \textit{Artificial Life Framework (ALF)} developed by Dalquen {\em et al.}~\cite{dal:ani:gon:des:11}. 
\textit{ALF} evolves gene sequences based on a probabilistic model 
along a randomly generated tree, starting with an ancestral gene. 
During this process evolutionary events are logged such that the \textit{true} MSA is known for each simulated gene family. This \textit{true} MSA can then be used as reference to assess the quality of automatically generated alignments.
 
We generated a series of 14 data sets, each containing 30 simulated
`genomes', with increasing mutation rates for the different data 
sets. For all other parameters in {\em ALF}, we used the default settings. 
In each data set, there are 750 simulated gene families such that one gene
from each gene family is present in each of the 30 simulated genomes. 
Thus, each of the `genomes' contains the same set of 750 genes. 
We varied the mutation rates between an average of 0.1013 substitutions per position for the first data set to an average of 0.8349 substitutions per position for the 14th data set.
The maximal pairwise distances between all pairs of sequences within one data set ranges from 0.1640 for the first to 1.0923 for the 14th data set. The 
simulated genes have an average length of about $1500 bp$, summing up to a 
total size of about 32 {\em MB} per data set.

To assess the quality of the produced alignments, 
we calculated \textit{recall} and \textit{precision} values in the usual way.
If, for one given data set, $S$ is the set of all positions in the 30 simulated 
genomes, we denote by 
$A\subset S^2$ the set of all pairs of positions aligned by the alignment that is to be evaluated while $R\subset S^2$ denotes the set of all pairs of 
positions aligned
in the reference alignment. \textit{recall} and \textit{precision} are then 
defined as  
\begin{align} 
recall=\frac{|A\cap R|}{|R|}, \qquad precision=\frac{|A\cap R|}{|A|}
\end{align}
The harmonic mean of {\em reall} and {\em precision} is called the {\em balanced F-score} and is often used as an overall  measure of  accuracy; it is thus 
defined as
$$ F_{score}=2\times \frac{precision\times recall}{precision + recall}$$
To estimate these three values, we used the tool \textit{mafComparator} which was also used in the \textit{Alignathon} study \cite{ear:ngu:hic:etal:14}. 
Since it is impractical to consider the entire set $S^2$ of pairs of positions of the test sequences, we sampled 10 million pairs of positions for each data set. This corresponds to the evaluation procedure used in  \textit{Alignathon}.

\begin{figure}[h!] 
\hspace{4mm}
\begin{center}
\includegraphics[width=12cm]{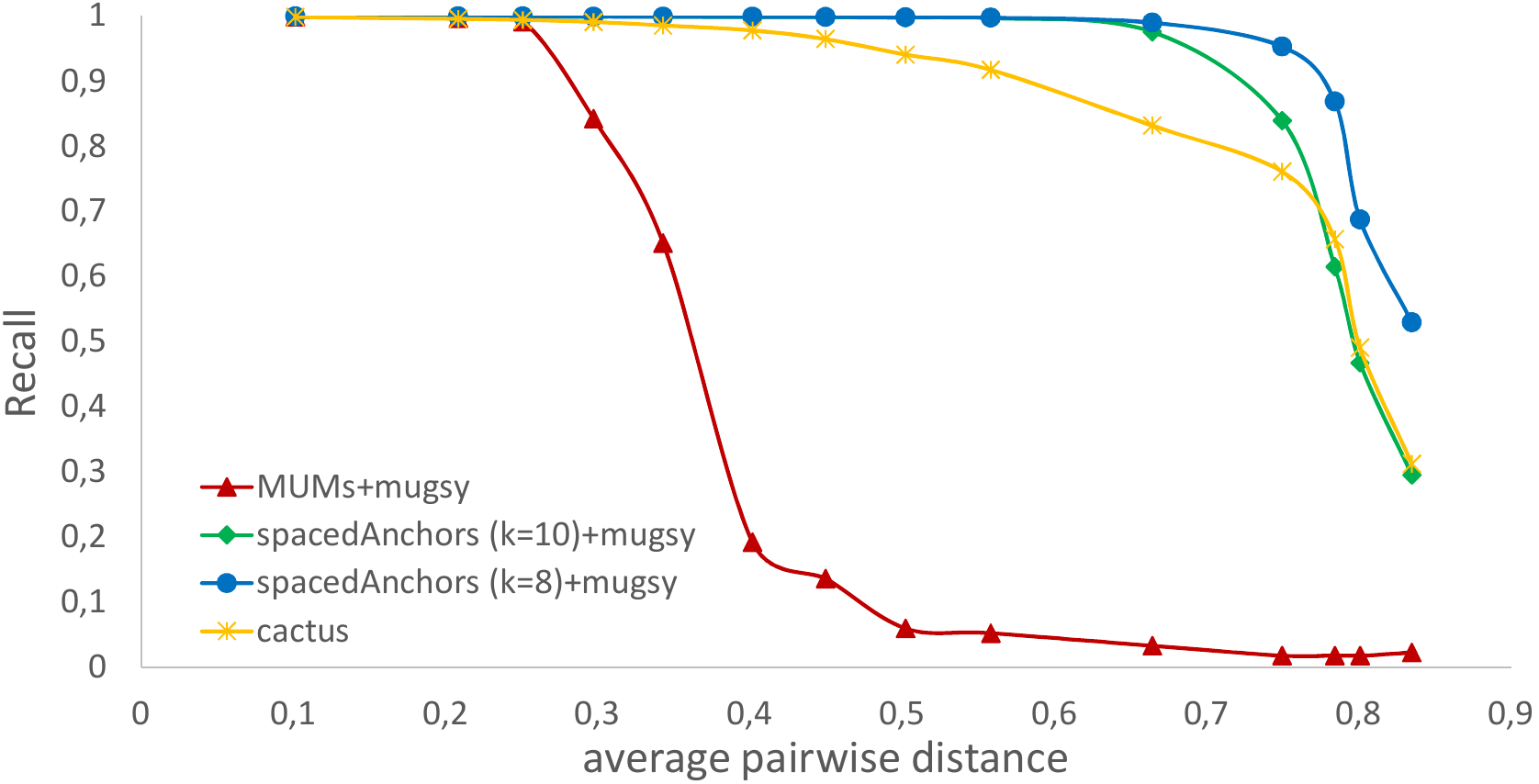}
\includegraphics[width=12cm]{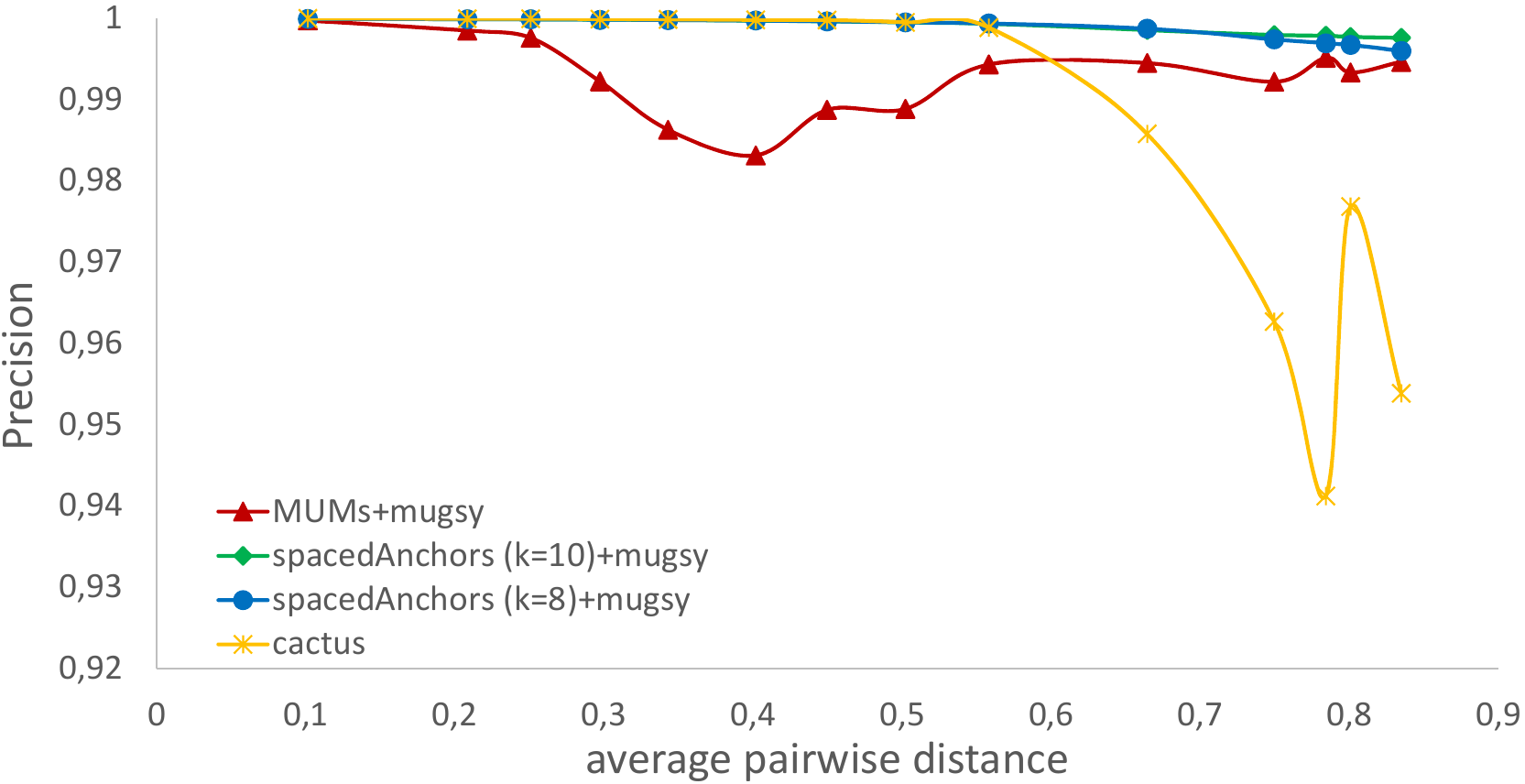}
\end{center}
\caption{{\em Recall} and {\em Precision} of \textit{Mugsy} with anchor points 
from \textit{Filtered Spaced Word Matches (FSWM)} and \textit{MUMmer}, 
respectively, and of \textit{Cactus} on simulated genomic sequences generated 
with \textit{ALF}, see main text for details. {\em FSWM} was used
with the default {\em weight} $w=10$, {\em i.e.} with
10 {\em match positions} in the underlying pattern. 
In addition, we ran {\em FSWM} with $w=8$.
\label{fig_alf_1}} 
\end{figure}

For the simulated sequence sets, their {\em precision} and {\em recall} values 
are shown in Figure~\ref{fig_alf_1}. 
For data sets with smaller mutation rates, 
alignments obtained with \textit{FSWM} are only slightly better than those obtained with \textit{MUMmer}. However, if the mutation rate increases, our spaced-words approach substantially 
outperforms the original version of \textit{Mugsy} where exact word matches
are used to find anchor points. 
Not only more homologies are detected but also the \textit{precision} is slightly higher if \textit{Filtered Spaced Word Matches} is used instead of {\em MUMmer}.

\begin{figure}[h!]
\hspace{4mm}
\begin{center}
\includegraphics[scale=\figscale]{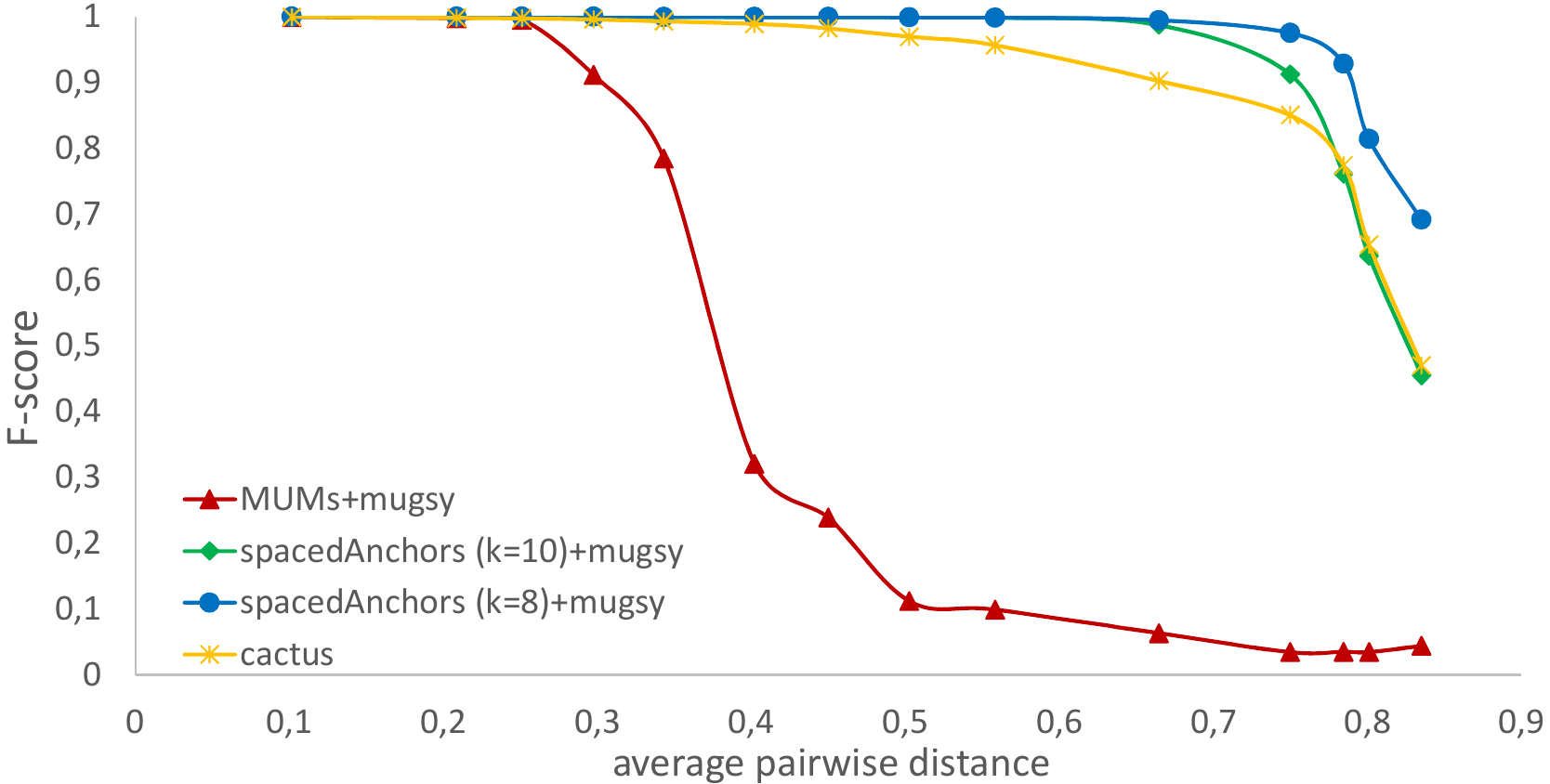}
\end{center}
\caption{{\em F-Score}  of \textit{Mugsy} with anchor points from \textit{Filtered Spaced Word Matches} and \textit{MUMmer}, respectively, and of \textit{Cactus} on simulated genomic sequences generated with \textit{ALF}.\label{fig_alf_2}}  
\end{figure}

 \subsection*{Real-world genome sequences }

For real-world genome families, it is usually not possible to calculate the {\em precision}
of MSA programs because it is, in general, not known which sequence positions exactly are
homologous to each other and which ones are not. 
If there are {\em core blocks} of the sequences for which
the biologically correct alignment is known, at least the {\em recall} can be
calculated for these core blocks. For most genome sequences, however, no such
core blocks are available.   
To evaluate {\em Mugsy}, the authors of the program used the 
number of {\em core columns} of the produced alignments as a criterion  
for alignment quality \cite{ang:sal:11}. Here, a {\em core column} is defined as
a column that does not contain gaps, {\em i.e.} a column that aligns 
nucleotides from all of the input sequences.
In addition, the authors of {\em  Mugsy} used the {\em number of pairs of aligned  
positions} of the aligned sequences as an indicator of alignment quality.  
In this paper, we are using the same criteria to evaluate multiple alignments of
real-world genomes.

As a first real-word example, we used a set of  29 {\em E.coli/Shigella} 
genomes that has already been used in the original {\em Mugsy} paper, 
see {\em supplementary material} for details; 
 these sequences have  also been used to evaluate
alignment-free methods \cite{hau:klo:pfa:14,yi:jin:13,mor:zhu:hor:lei:15}. 
The total size of this data set is about 141 {\em MB}.
As a second test set, we used another prokaryotic data set which consists of 32 complete \textit{Roseobacter} genomes (details in the {\em supplementary material}). 
This data set was used to assess the performance on more distantly related organisms than the {\em E.coli/Shigella} strains. The total size of these data set is about 135 {\em MB}.  
To test our approach on eukaryotic genomes, we used 
as a third test case a set of nine fungal genomes, namely  
 \textit{Coprinopsis cinerea}, 
    \textit{Neurospora crassa}, 
    \textit{Aspergillus terreus}, 
    \textit{Aspergillus nidulans}, 
    \textit{Histoplasma capsulatum}, 
    \textit{Paracoccidioides brasiliensis}, 
    \textit{Saccharomyces cerevisiae}, 
    \textit{Schizosaccharomyces pombe} and  
    \textit{Ustilago maydis} 
(genbank accession numbers are given in the supplementary material).  
The total size of this third data set is about 253 {\em MB}.
The results of {\em Mugsy} with {\em MUMmer} and {\em FSWM} for the three real-world data sets are shown in Table~1, 
together with the results obtained with {\em Cactus}. In addition to the number of {\em core columns} and the number of aligned pairs of positions,
the table contains 
the number of {\em core Locally Collinear Blocks}, {\em i.e.} the number of {\em Locally Collinear Blocks} involving all of the input sequences, 
and the total number of {\em Locally Collinear Blocks}  
returned by the alignment programs.

\begin{center}
\begin{table}[h!]
\begin{center}
    \begin{tabular}{| l | r | r | r | r |}
    \hline
     & $\#$ core LCBs & $\#$ aligned pairs & $\#$ core col. & $\#$ LCBs\\
        \hline
        \multicolumn{5}{|c|}{29 {\em E.coli/Shigella} genomes}\\
        \hline
        \textit{Mugsy}  + \em{MUMmer}
        & 539 & 1,61E+09 & 2,827,115 & 4,138 \\
        \hline
        \textit{Mugsy} + {\em FSWM}
        & 664  & 1,63E+09 & 2,867,432  & 5,906 \\
        \hline
        \textit{Cactus}
        &20,163  & 1,48E+09 & 2,663,750  & 56,592 \\
        \hline
        \multicolumn{5}{|c|}{ 32 {\em Roseobacter} genomes }\\
        \hline

        \textit{Mugsy} + {\em MUMmer} & 39   & 3,63E+08 & 13,654  & 13,501  \\
\hline
        \textit{Mugsy} + {\em FSWM}
& {859}  &  {7,15E+08} & {824,054} & {30,836} \\
        \hline
        \textit{Cactus} &5,984  & 4,95E+08 & 280,085   & 337,320 \\
        \hline

        \multicolumn{5}{|c|}{ 9 fungal genomes }\\

        \hline
        \textit{Mugsy} + {\em MUMmer} & 9 & 5,88E+06 & 2,097 & 4,252 \\
        \hline
        \textit{Mugsy} + {\em FSWM} & {2,590} & {1,18E+08} & {718,176} & {89,555} \\
        \hline
        \textit{Cactus} & 31,589 & 1,33E+08 & 828,680 & 848,242 \\
        \hline

    \end{tabular}
\end{center}
\vspace{3mm}

    \caption{Multiple alignments of 29 {\em E.coli/Shigella} genomes,
32 {\em Roseobacter} genomes and 9 fungal genomes,
 calculated
with \textit{Mugsy} using anchor points from our \textit{spaced-words} approach
and from \textit{MUMs}, respectively, and with {\em Cactus}.
The first column contains the number of {\em core columns}, {\em i.e.} the number of columns in the
multiple alignment that do not contain gaps; the second column contains the total number of aligned pairs of
positions in the alignment. The third column contains the number of {\em core Locally Collinear Blocks (LCBs)}
{\em i.e.} the number of {\em LCBs} that involve {\em all} of the aligned genomes (`core LCBs'),
while the last column contains the total number of {\em LCBs}.
\label{real_genomes}
}
\end{table}
\end{center}

\subsection*{Program run time}

Table~2 reports 
the program run times of {\em Mugsy} with {\em FSWM}, {\em Mugsy}
with {\em MUMmer} and {\em Cactus}  
on the above three real-world sequence sets.  
In addition, the table contains the run times for 
{\em FSWM} and {\em MUMmer} alone.

\begin{center}
\begin{table}[h]
    \begin{tabular}{| l | c | c | c |}
    \hline
     & \textit{E.coli/Shigella} & \textit{Roseobacter} &\textit{fungal genomes}\\
        \hline
        \textit{FSWM} & 59 & 83 & 110 \\
        {\em FSWM} + {\em Mugsy}& 638 &6428 & 1488 \\
        \hline
        \textit{MUMmer} & 73 & 63 &  43 \\
        {\em MUMmer} + {\em Mugsy} & 286 & 1099 & 63  \\
        \hline
        \textit{Cactus} & 714 & 1775 & 775  \\
        \hline
    \end{tabular}
     \caption{Run time in minutes for three different 
multiple genome-alignment methods applied to the three test data sets that
we used in our program evaluation. }\label{time}
\end{table}
\end{center}

\section{Discussion}
In this paper, we proposed a novel approach to calculate anchor points for genome alignment.
Finding suitable anchor points is  a critical step in all methods for genome 
alignment, since the
selected anchor points determine which regions of the sequences can be aligned
to each other in the final alignment. 
A sufficient number of anchor points is necessary to keep the  
search space and run time of the main alignment procedure manageable, so 
{\em sensitive} methods are needed to find anchor points. 
Wrongly selected anchor points, on the other hand, can seriously deteriorate 
the quality of the final alignments, so anchoring procedures 
must also be highly {\em specific}.  

Earlier approaches to genomic alignment used exact word matches as anchor points \cite{del:kas:fle:99,hoe:kur:ohl:02}, 
since such matches can be easily found using suffix trees and related 
indexing structures. These approaches are limited, however,
to situations where closely related   genomes are to be aligned, 
for example different strains of a bacterium. 
In modern approaches to database searching, {\em spaced seeds} are used 
to find potential sequence homologies
  \cite{li:ma:kis:tro:03,hau:sin:rei:14,buc:xie:hus:15}. 
Here, binary patterns of {\em match} and {\em don't care} positions are used, 
and two sequence segments of the corresponding length
are considered to match if identical 
residues are aligned at the {\em match} positions, while mismatches are 
allowed at the {\em don't care}  positions. 
Such pattern-based approaches are more {\em sensitive} than previous methods 
that relied on exact word matches.

We previously proposed to apply the `spaced-seeds' idea to alignment-free 
sequence comparison, by replacing contiguous words by  
so-called {\em spaced words}, {\em i.e.} by words that contain wildcard
characters at certain pre-defined positions \cite{lei:bod:hor:lin:mor:14}.  
More recently, we introduced {\em filtered spaced word matches}
\cite{lei:soh:mor:17} to estimate phylogenetic distances between genome
sequences. 
In the latter approach, we first identify spaced-word matches 
using relatively long patterns with only few {\em match} positions. 
For the identified
matching segments, we then look at {\em all} aligned  pairs of nucleotides, 
including the ones at the {\em don't-care} positions, and we discard 
spaced-word matches if the overall degree of similarity between the two 
segments is below a threshold. 
Phylogenetic distances can be estimated based on the aligned nucleotides at 
the don't-care positions of the remaining spaced-word matches.  
We showed that this procedure is fast and highly sensitive, and it can 
reliably distinguish between true homologies  and spurious sequence
similarities.     

In the present study, we used filtered spaced word matches to calculate 
high-quality anchor points  for genomic sequence alignment. 
Instead of using spaced-word matches directly as anchor points, we extend
them into both directions, similar to the {\em hit-and-extend} approach to database 
searching. To evaluate these anchor points, we integrated them into the
popular genome-alignment pipeline {\em Mugsy}. 
Test runs on simulated genome sequences 
show that, for closely related sequences,  {\em Mugsy} produces alignments 
of high quality with both types of anchor points. 
For more distantly related sequences, however, the {\em recall} values
of the program drop dramatically if anchor points are calculated with
{\em MUMmer} while, with our spaced-word matches, one observes
recall values close to 100$\%$ for distances up to around 0.7 substitutions 
per position.  

For real-world genomes, it is more difficult to evaluate the performance of genome aligners since
there is only limited information available on which positions are homologous to each other and which ones are not. 
Angiuoli and Salzberg \cite{ang:sal:11} 
therefore used the number of aligned pairs of positions as an indicator of 
alignment quality, together with the size of the `core alignment', {\em i.e.} the number of 
alignments columns  that do not contain gaps. At first glance, 
these criteria might seem questionable; it would be trivial to maximize these 
values, simply by aligning sequences without internal gaps, by adding gaps
only at the ends of the shorter sequences. However, as shown in 
Figure~\ref{fig_alf_1}, all MSA programs in our study have high 
{\em precision} values, {\em i.e.} positions aligned by these programs
are likely to be true homologs. In this situation, the number of aligned 
position pairs and size of the `core alignment' can be considered as a proxy 
for the {\em recall} of the applied methods {\em i.e.} the proportion of
homologies that are correctly aligned.

For distantly related sequence sets, the total run time of {\em Mugsy} 
is much higher with our {\em FSWM} anchoring approach than with {\em MUMmer}.  
One reason for the increased run time with {\em FSWM}
is the fact that, with spaced-words, far more 
{\em Locally Collinear Blocks} are detected, than if exact word matches
are used as anchor points,
 especially for distantly related sequences where exact word matching is 
not very sensitive. 
One possible solution for this issue would be to apply 
user-defined threshold values for the total number of returned
{\em Locally Collinear Blocks} or for their similarity scores, to reduce
the run time of the final alignment procedure 
for large  genomic sequences.



\bibliographystyle{abbrv}

\bibliography{/home/bmorgen/my_papers/bibtex/all_papers}

\end{document}